# Disabled Access for Museum Websites

Jonathan P. Bowen

London South Bank University
Faculty of BICM, Borough Road
London SE1 0AA, UK
+44 (0)20 7815 7462

jonathan.bowen@lsbu.ac.uk
www.jpbowen.com

## ABSTRACT
Physical disabled access is something that most museums consider very seriously. Indeed, there are normally legal requirements to do so. However, online disabled access is still a relatively novel field. Most museums have not yet considered the issues in depth. The Human-Computer Interface for their websites is normally tested with major browsers, but not with specialist browsers or against the relevant accessibility and validation standards. We consider the current state of the art in this area and mention an accessibility survey of some museum websites.

## Categories and Subject Descriptors
H.5.1 [**Information interfaces and presentation (e.g., HCI)**]: User Interfaces – *benchmarking, evaluation/methodology, graphical user interfaces (GUI), input devices and strategies, interaction styles, screen design, standardization, style guides, user-centered design, voice I/O.*

## Keywords
Accessibility, Disabled Access, Museums, Usability, WWW.

## 1. INTRODUCTION
A cartoon in *The New Yorker* magazine on 5 July 1993 showed a picture of a dog at a computer talking to his friend and saying:

> *On the Internet, nobody knows you're a dog.*
> – Peter Steiner

A significant number of people using the Internet in general and the web in particular have some form of disability that may affect their use of the technology. Of course the World Wide Web Consortium (W3C) is aware of the issues [9] but many sectors still have little awareness of the access problems to their websites. Museums normally pride themselves on the accessibility of their physical buildings but many have yet to make an equivalent effort for their online facilities, despite the fact that legislation is in the offing or already exists in most developed countries.

## 2. EXAMPLE MUSEUM WEBSITES
It is instructive to consider some museum sites where accessibility has been considered. This is now an aspect that is assessed in the *Museums and the Web* conference *Best of the Web awards*

[www.archimuse.com/mw2003/best]. As an example, the Natural History Museum of Los Angeles County in the US [www.nhm.org] has been an exemplary website from the point of view of accessibility. This was largely because someone who is very knowledgeable of the issues involved designed and organized it in-house over a number of years. Perhaps the most interesting thing about this and other websites designed with accessibility in mind is that they need look no different on a modern graphical web browser from any other professionally designed website. This demonstrates that designing with accessibility in mind does not mean one has to compromise what is on offer for the able bodied user with good web browsing facilities.

The British Museum COMPASS database of a selection of the museum's best objects [www.thebritishmuseum.ac.uk/compass] includes a prominent 'TEXT ONLY' link at the top of its homepage for disabled users to gain easy access to the available facilities. The information for the graphical and text-based information is served from the same database, thus ensuring that both are in step and up to date.

The Tate Gallery in London initiated the i-Map Project in associated with a major exhibition on Matisse and Picasso in 2002 [www.tate.org.uk/imap]. This was designed to give access for visually impaired people via the web using raised images allowing them to be touched if printed on a special printer. Thus even art galleries that are obviously very visually oriented in general can make efforts to reach out to the blind.

## 3. SURVEY RESULTS
The Bobby validator [bobby.watchfire.com], which can check for WAI compliance [9] and also the US Government Section 508 compliance [www.section508.gov] has been used to evaluate the accessibility and usability of 25 UK and 25 international museum and related websites with respect to their disabled accessibility (and hence usability) [3,5,6]. Bobby evaluates web pages for accessibility to users with disabilities. It checks for the presence or absence of particular features, or their characteristics, although it does not explicitly check HTML syntax. The W3C validator is recommended for checking HTML itself [validator.w3.org]. The results of this survey can be found in [6].

As well as the mechanical check using Bobby, a visual analysis was carried out manually using information provided by Bobby and also with textual browsing (e.g., using an audio browser) and partial sightedness in mind. Generally the sites faired quite badly with a significant number exhibiting some serious accessibility problems. The overall results for Bobby are shown in Table 1, extracted from [6]. Priority 1 errors are most serious and must be corrected to meet the WAI guidelines [9]. Priority 2 errors should be corrected if possible and Priority 3 errors may be corrected.

Table 1. Bobby validation results for museum sites

| Bobby Validation | UK | % | International | % |
|---|---|---|---|---|
| Priority 1 | | | | |
| No errors | 10 | 40 % | 7 | 28 % |
| 1 error | 12 | 48 % | 16 | 64 % |
| 2-5 errors | 1 | 4 % | 2 | 8 % |
| Triggered items | 20 | 80 % | 19 | 76 % |
| Non-triggered items | 25 | 100 % | 25 | 100 % |
| Priority 2 | | | | |
| No errors | 1 | 4 % | 5 | 20 % |
| 1 error | 5 | 20 % | 5 | 20 % |
| 2-5 errors | 19 | 76 % | 17 | 68 % |
| Triggered items | 25 | 100 % | 24 | 96 % |
| Non-triggered items | 25 | 100 % | 25 | 100 % |
| Priority 3 | | | | |
| No errors | 2 | 8 % | 2 | 8 % |
| 1 error | 5 | 20 % | 8 | 32 % |
| 2-5 errors | 17 | 68 % | 16 | 64 % |
| Triggered items | 25 | 100 % | 25 | 100 % |
| Non-triggered items | 25 | 100 % | 25 | 100 % |

## 4. OTHER SOURCES OF INFORMATION

An excellent guide with museums specifically in mind is the Ed-Resources.Net Universal Access website by Jim Angus [www.ed-resources.net/universalaccess]. This includes an interesting comparison of the accessibility of three museum websites, as well as links to online web page validation services and further relevant resources.

In the UK, MAGDA, the Museums & Galleries Disability Association [www.magda.org.uk] is dedicated to improving access to UK museums and galleries for people with disabilities as well as disseminating current best practice. It also provides an online forum for museum and gallery professionals to enable online discussion [groups.yahoo.com/group/magdamail].

The number of books explicitly covering web accessibility has been very limited until recently, but at least three are now available. [7] was the first book in the area known to the author. [8] is written by a range of experts, including legal aspects in some detail for example, and as such is perhaps the most authoritative book in this area to date. [4], the most recent book, may be more approachable for some web designers. It is hoped that more books in this important subject area will be produced in the future.

## 5. CONCLUSION

This poster is intended to help raise awareness of the issues concerning disabled access online, especially in the context of museums that are increasingly developing their online resources with ever-more sophisticated web technologies.

For people with learning disabilities, visual disabilities, and reading impairments, print-based text can be completely inaccessible. While in recent years software developers have created electronic screen readers that convert text to speech, few of these programs offer effective control over how the text is displayed and read, nor do they provide flexible reading features. Therefore, for those with visual impairments, learning disabilities, reading disabilities, or language proficiency problems, even electronic text can be difficult to decipher. The World Wide Web poses additional barriers; while the web provides a great deal of useful, educational information, its reading levels, page design, and emphasis on graphics can make it inaccessible or unusable for some.

There is much room for improvement in the reading and speaking qualities of screen readers. The museum website accessibility survey mentioned here [6] has shown that in order to develop better accessibility the emphasis must be on improved web page coding practice. The coding aspects of web pages is extremely important in ensuring wide accessibility of websites that will be useful to all, both able bodied and disabled alike. This is possible with care and thought, but most web design professionals have yet to attain the skills to do this. It is hoped that this paper will at least raise some awareness and interest in the issues involved, particularly for museums and other public bodies that pride themselves in their accessibility.

The Museophile initiative [www.museophile.com], a spinout from South Bank University, aims to help museums online in areas such as e-commerce, discussion forums and accessibility. In particular, for online information on web accessibility for museums, see:

http://access.museophile.net

An expanded paper including fuller details is also available [3].

## 6. ACKNOWLEDGMENTS
Giuseppe Micheloni undertook the survey reported here as a final year project at South Bank University [5].

## 7. REFERENCES
[1] Bowen, J.P. Internet: A question of access. *New Heritage*, 04.01 (2001), 58.

[2] Bowen, J.P. Tackling web design & Advice on accessible website design. Museums Journal, 101, 9 (September 2001), 41-43.

[3] Bowen, J.P. and Micheloni, G. *Disabled Access for Museum Websites.* Technical Report, SCISM, South Bank University, London, UK, 2002. Presented at MCN2002.
http://www.museophile.lsbu.ac.uk/access/mcn2002/access.pdf

[4] Clark, J. *Building Accessible Websites*. New Riders, 2003.

[5] Micheloni, G. *An Accessible and Usable Art Gallery for All*. Final year project, SCISM, South Bank University, London, UK, 2002.

[6] Micheloni, G. and Bowen, J.P. *Accessibility and Usability Survey on UK and International Art Gallery and Museum Websites*. Technical Report, SCISM, South Bank University, London, UK, 2002.
http://www.museophile.lsbu.ac.uk/access/mcn2002/survey.pdf

[7] Paciello, M.G. *Web Accessibility for People with Disabilities*. CMP Books, 2000.

[8] Thatcher, J. et al. Constructing *Accessible Web Sites*. Glasshaus, 2002.

[9] W3C. *Web Accessibility Initiative (WAI)*. 1994-2003.
http://www.w3.org/WAI/